# Discovery of a superhard iron tetraboride superconductor

Huiyang Gou[1,2], Natalia Dubrovinskaia[2*], Elena Bykova[1,2], Alexander A. Tsirlin[3], Deepa Kasinathan[3], Asta Richter[4], Marco Merlini[5], Michael Hanfland[6], Artem M. Abakumov[7], Dmitry Batuk[7], Gustaaf Van Tendeloo[7], Yoichi Nakajima[1], Aleksey N. Kolmogorov[8], Leonid Dubrovinsky[1]

[1] *Bayerisches Geoinstitut, Universität Bayreuth, D-95440 Bayreuth, Germany*

[2] *Material Physics and Technology at Extreme Conditions, Laboratory of Crystallography, University of Bayreuth, D-95440 Bayreuth, Germany*

[3] *Max Planck Institute for Chemical Physics of Solids, Nöthnitzer Str. 40, D-01187 Dresden, Germany*

[4] *Technische Hochschule Wildau, Bahnhofstrasse 1, D-15745 Wildau, Germany*

[5] *Dipartimento di Scienze della Terra, Università degli Studi di Milano, Via Botticelli 23, 20133 Milano, Italy*

[6] *ESRF, Boîte Postale 220, 38043 Grenoble, France*

[7] *EMAT, University of Antwerp, Groenenborgerlaan 171, B-2020 Antwerp, Belgium*

[8] *Department of Physics, Applied Physics and Astronomy, Binghamton University, State University of New York, Vestal, NY 13850, USA*

*natalia.dubrovinskaia@uni-bayreuth.de

**Single crystals of novel orthorhombic (space group *Pnnm*) iron tetraboride $FeB_4$ were synthesised at pressures above 8 GPa and high temperatures. Magnetic susceptibility measurements demonstrated bulk superconductivity below 2.9 K. The putative isotope effect on the superconducting critical temperature indicates that $FeB_4$ is likely a phonon-mediated superconductor, which is unexpected in the light of previous knowledge on Fe-based superconductors. The discovered iron tetraboride is highly incompressible and has the nanoindentation hardness of 65(5) GPa, thus, it opens a new class of highly desirable materials combining advanced mechanical properties and superconductivity.**

Modern computational materials design is gaining broad recognition as an effective means of reducing the number of experiments that can ultimately lead to materials discovery [1-3]: successful examples now include Li-ion batteries, thermoelectrics, cathode materials, catalysts, etc. Superconductors remain one of the most challenging classes of materials to develop [2,4-6]. There have been strikingly few cases though where theory successfully guided the experiment to a discovery and those were essentially restricted to thoroughly study elemental materials, namely, silicon [7] and lithium [8] under pressure. The progress can be attributed to the improvement of density functional theory (DFT)-based methods [9,10], advances in compound prediction strategies [1,3], and the steady growth of computational resources. Still there are number of problems in predictions of novel superconductors [4]. Firstly, only one type of superconductors, called phonon-mediated or conventional, is understood well enough [4] to be described by theories with predictive power [5,11]. Calculation of the superconducting critical temperature, $T_c$, is possible but exceedingly demanding as a viable option in high-throughput screening for candidate materials. Secondly, the inverse correlation between the stability of a compound and its phonon-mediated superconducting $T_c$ has been pointed out in a number of studies: a considerable density of states (DOS) at the Fermi level, beneficial for high $T_c$, is often an indication of structural instability [6]. One of the remarkable exceptions is the stoichiometric $MgB_2$ material [12] with naturally hole-doped σ$f$ bands and a $T_c$ of 39 K.

The problem of thermodynamic instability can be mitigated under high pressure. When quenched to normal conditions, materials with the large DOS at the Fermi level may remain metastable and show superconductivity facilitated by this large DOS. Kolmogorov et al. [9] systematically examined the Fe-B system and showed that a previously unknown compound, $FeB_4$, may exist under normal conditions in a never-seen-before orthorhombic crystal structure. The material was predicted to have naturally electron-doped bands and a large electron-phonon coupling that might render $FeB_4$ the first *conventional* Fe-based superconductor [9], as opposed to the recently discovered family of *unconventional* Fe-based superconductors [2, 13]. Bialon et al. [14] suggested that the predicted $FeB_4$ phase could be synthesised under pressure. Wide and growing interest in studies of Fe-based superconductors [2], simple chemical composition, and expected mild pressure-temperature conditions for synthesis [14] make iron tetraboride a good case for testing the computational predictive power and, thus, the degree of our theoretical comprehension of such a complex physical phenomenon as superconductivity. Here, for the first time, we verify experimentally the computational prediction of a novel superconductor.

The experimental Fe-B phase diagram [15] at ambient pressure is very poor in compounds. So far only represented by tetragonal $Fe_2B$ and orthorhombic FeB (Ref. [16]), although hexagonal $FeB_2$ (Ref. [17]) and rhombohedral $FeB_{\sim 49}$ (Ref. [18]) have also been reported in literature. Additionally to the earlier calculated orthorhombic $Fe_3B$ phase [19], recently two new orthorhombic phases were theoretically predicted in the Fe-B system [9], oP12- $FeB_2$ prototype as the ground state for $FeB_2$ and $FeB_4$.

We have undertaken a series of high pressure experiments [20] aimed at synthesis of the predicted Fe-B phases (Table S1, Ref. [20]). Independent on pressure, a major component of the reacted mixture was stoichiometric FeB. At low pressures (3 GPa and below) and temperatures of 1323 K to 1973 K only known phases, orthorhombic FeB and rhombohedral $FeB_{\sim 49}$, were produced. Experiments at pressures of 8 GPa to 18 GPa and temperatures of 1523 K to 2023 K (Table S1) led to the synthesis of previously unidentified orthorhombic $FeB_4$, $Fe_2B_7$, and tetragonal $Fe_{1+x}B_{50}$ ($x \approx 0.04$) phases. The compounds crystallize from the melt and by optimizing the sample geometry, heating duration, and temperature gradients along the capsules it was possible to increase the amount of boron-rich Fe-B phases. However, as seen in Fig. 1*a*, all the products of the high-pressure synthesis, and particularly $FeB_4$ and $Fe_2B_7$, are found in a tight mutual intergrowth, so that the procedure of phase separation is challenging. The largest pieces of pure $FeB_4$ produced so far have dimensions on the order of 150x150x100 $\mu m^3$.

The chemical composition and purity of the newly synthesised phases were revealed by the microprobe wavelength dispersive X-ray (WDX) and energy dispersive X-ray (EDX) analysis (performed in SEM and TEM) [20]. The crystal structures of $FeB_4$, $Fe_2B_7$, and $Fe_{1+x}B_{50}$ have been solved from single crystal X-ray diffraction data (Table 1). A detailed description of $Fe_2B_7$ and $Fe_{1+x}B_{50}$ is out of the scope of the present paper and will be published elsewhere.

According to the single crystal X-ray and electron diffraction [20], $FeB_4$ adopts an orthorhombic *Pnnm* (Z = 2) crystal structure. The refined structure was confirmed by high angle annular dark field scanning transmission electron microscopy (HAADF-STEM) images along the [100], [010] and [001] directions (Fig. 1*b*, Figs. S9, S10). Additionally, planar defects confined to the (010) planes were occasionally observed in $FeB_4$. These defects are not abundant in the material as indicated by the absence of any related diffuse intensity on the electron diffraction patterns (Fig. S8).

A polyhedral model of the FeB$_4$ structure is shown in Fig. S1 (Ref. [20]). The structure is remarkably close to that theoretically predicted [9] (Table 1), and suggested also for CrB$_4$ based on the same first-principle calculations [21].

Despite the very small size of the available phase-pure samples, we were able to confirm the prediction of superconductivity in FeB$_4$. While resistivity measurements are presently unfeasible, magnetic susceptibility data unequivocally demonstrate bulk superconductivity in FeB$_4$. Magnetic susceptibility measurements under zero-field-cooling (ZFC) conditions reveal a strong diamagnetic response of FeB$_4$ samples below 3 K (Fig 2). Above 3 K, FeB$_4$ is weakly paramagnetic with an additional ferromagnetic signal emerging below 30 K (Figs. S3-S7, Ref. [20]). The strong diamagnetic response of FeB$_4$ is a clear footprint of bulk superconductivity. The drop in the volume susceptibility ($\chi_V$) is $4\pi(\Delta\chi_V) = -1.3$ and slightly exceeds the value of $4\pi\chi_V = -1$ expected for an ideal superconductor [22]. A magnetic field suppresses the $T_c$ and eventually destroys superconductivity above 100 mT. To elucidate the nature of the observed superconducting transition, we compared the transition temperatures in the samples containing different boron isotopes (Fig. 3). The sample enriched with the heavier B isotope shows a lower $T_c$ (2.95(1) K and 2.89(1) K for the $^{10}$B and $^{11}$B samples, respectively), as expected for a phonon-mediated superconductor, where phonon frequencies change with the atomic mass. Indeed, our tentative estimate of the isotope effect (Supplementary Information) yields $\Delta T_c \sim 0.05$ K in good agreement with $\Delta T_c \sim 0.06(2)$ K, as found experimentally. Note that the difference in the theoretically predicted (~20 K (Ref. [9])), and experimentally observed $T_c$ can be related to the imperfectness of real crystals revealed by HAADF-STEM. Above 3 K, the FeB$_4$ samples show weak ferromagnetism with very low values of the magnetic moment on the order of 0.01 $\mu_B$/f.u. This effect is probably extrinsic (Fig. S5) because we cannot fully exclude the presence of micro quantities (not detectable even by synchrotron X-ray diffraction) of crystalline [9] or amorphous [23] strongly ferromagnetic impurities of Fe$_{1-x}$B$_x$ compound(s). Above 30 K, FeB$_4$ is a conventional Pauli paramagnet with a nearly temperature-independent magnetic susceptibility.

Metal borides are known for their high hardness [24], so that characterisation of the elastic behavior of the newly synthesized boride and its stability under pressure is an important issue. No phase transitions were observed under compression of FeB$_4$ at ambient temperature in a diamond anvil cell up to ca. 40 GPa (Ref. [20]). Compressibility measurements on both compression and decompression revealed the relatively high bulk modulus, $K = 252(5)$ GPa ($K' = 3.5(3)$, $V_0 = 72.79(4)$ Å) (Fig. 3$a$), and a significant degree of anisotropy in the elastic behaviour of FeB$_4$. The structure of FeB$_4$ is most compressible along the *a*-direction, while stiffest along the *b*-axis (Fig. 3*b*). It may be related to the fact that the shortest (and thus least compressible) B-B contact (1.714(6) Å at ambient conditions) in this structure is almost parallel to the *b*-axis. The stiffness of the FeB$_4$ structure along the *b*-direction is the same as that of diamond [25] (Fig. 3*b*) suggesting that the iron tetraboride may have remarkably advanced mechanical properties. Figure 3 *c,d* presents the results, which are obtained by an average over several nanoindentation load-displacement charts on FeB$_4$ without the feature of a pop-in [20]. The depth dependent indentation or reduced modulus E$_r$ shows a clear plateau with E$_r$ = 633±30 GPa (Fig. 3*c*) that is far ahead compared to common ceramic materials like alumina [26] (~350 GPa) at room temperature. However, Young's moduli of diamond [27] (~1000 GPa) and cubic boron nitride [28] (~900 GPa) are still considerably larger. Nevertheless the nanoindentation hardness approaches an average value of H = 62±5 GPa (Fig. 3*d*). Microhardness measurements were difficult to conduct because of the small size of the phase-pure samples of FeB$_4$. However, several successful tests (Fig. S2) with a load of 20 N gave values of the Vickers hardness ranging from 43 to 70 GPa, thus confirming that FeB$_4$ belongs to the group of superhard materials [29].

Summarizing the results, it is worth underlining that before the present work was undertaken, (i) there was no compound known with the FeB$_4$ composition, (ii) the predicted orthorhombic crystal structure [9] was not yet observed in any material, (iii) the phonon-mediated superconductivity theoretically suggested for FeB$_4$ was not anticipated for the Fe-based materials previously known as unconventional superconductors [2,4]. In addition, the newly synthesized compound was found to be superhard, well exceeding the expectations about its potential mechanical properties [21]. This finding, bridging the gap between the superhardness and superconductivity community, may lead, for example, to a possibility for designing new superconducting nanoelectromechanical systems and/or observation of new fundamental effects.

Thus, the prediction [9] and consequent experimental proof of an 'unlikely phonon-mediated', superhard FeB$_4$ superconductor, reported in the present work, is a result of synergy of high-pressure and computational materials research. FeB$_4$ is the first material which in pure, single crystal form combines superconductivity and superhardness, and thus opens a new highly desirable class of materials.

**Acknowledgements** The work was supported by the German Research Foundation (DFG). N.D. thanks DFG for financial support through the Heisenberg Program and the DFG Project DU 954-8/1. H.G. gratefully acknowledges financial support of the Alexander von Humboldt Foundation. A.M.A., D.B. and G.V.T. acknowledge support from the ERC grant N°246791 "COUNTATOMS".

TABLE I. Experimental single crystal X-ray diffraction data for $FeB_4$ and the results of its structure solution compared to the structural data of $FeB_4$ theoretically predicted by Kolmogorov *et al.* [9]

| Empirical formula | $FeB_4$ | $FeB_4$ (Ref. [9]) |
|---|---|---|
| **Crystal system** | Orthorhombic | Orthorhombic |
| **Space group** | *Pnnm* | *Pnnm* |
| *a* (Å) | 4.5786(3) | 4.521 |
| *b* (Å) | 5.2981(3) | 5.284 |
| *c* (Å) | 2.9991(2) | 3.006 |
| *V* (Å$^3$) | 72.752(8) | 71.810 |
| *Z* | 2 | 2 |
| **Atomic coordinates** (*x/a, y/b, z/c*) | | |
| **Fe1, 2***a* | 0, 0, 0 | 0, 0, 0 |
| **B1, 4***g* | 0.2487(9), 0.3123(7), 0 | 0.2508, 0.3129, 0 |
| **B2, 4***g* | 0.3411(8), 0.1263(7), 1/2 | 0.3394, 0.1267, 1/2 |
| **Calculated density (g/cm$^3$)** | 4.523 | |
| **Crystal size (mm$^3$)** | 0.05x0.04x0.04 | |
| **Theta range for data collection (deg.)** | 5.89 to 36.13 | |
| **Completeness to theta = 25°, %** | 100 | |
| **Reflections collected** | 896 | |
| **Independent reflections / $R_{int}$** | 193 / 0.0345 | |
| **Data [*I* > 2σ(*I*)] / restraints / parameters** | 164 / 0 / 17 | |
| **Goodness of fit on *F*$^2$** | 1.094 | |
| **Final *R* indices [*I* > 2σ(*I*)]** | 0.0279 / 0.0615 | |
| *R$_1$*/*wR$_2$* | | |
| ***R* indices (all data)** | 0.0400 / 0.0666 | |
| *R$_1$*/*wR$_2$* | | |
| **Largest diff. peak and hole (e / Å$^3$)** | 0.924 and -1.090 | |

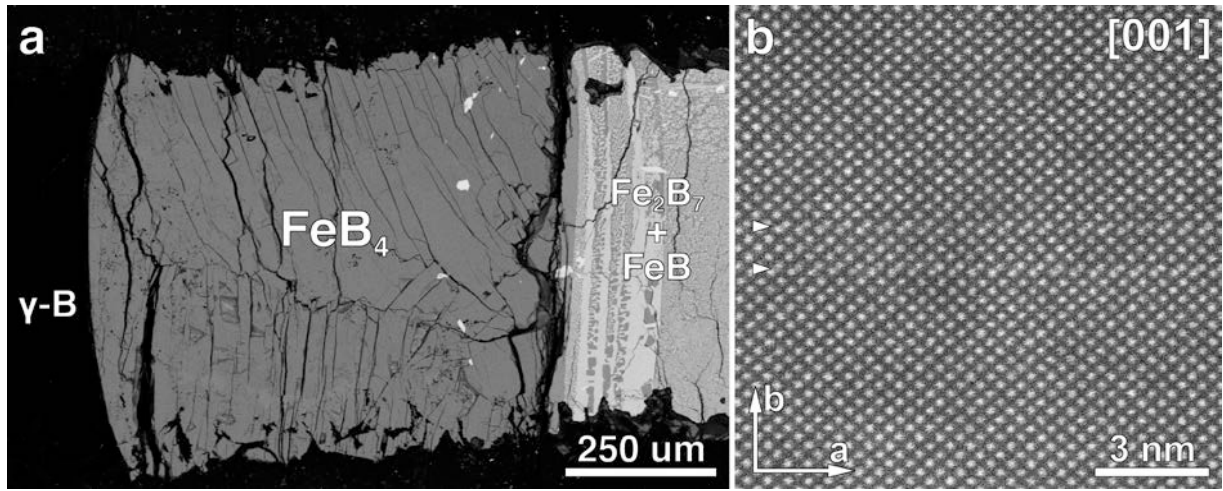

FIG. 1. (*a*) The backscattered electron SEM image of the polished surface of a high-pressure sample. The central part of the image (dark gray field) represents FeB$_4$ produced by the reaction of Fe with B after melting. The adjacent area on the right appears brighter as it is composed of the phases with lower boron content, namely Fe$_2$B$_7$ and FeB. The surrounding black field is non-reacted boron which, however, underwent a pressure-induced phase transformation from β-B to γ-B. Boron intrusions also fill the cracks in the FeB$_4$ phase. (*b*) The high resolution [001] HAADF-STEM image of FeB$_4$ (bright dots correspond to the Fe columns). Occasional planar defects (marked with arrowheads) are confined to the (010) plane and are visible as lines running parallel to the *a*-axis and consisting of pairs of the Fe columns with a shorter projected intercolumn distance in comparison with the FeB$_4$ matrix [20].

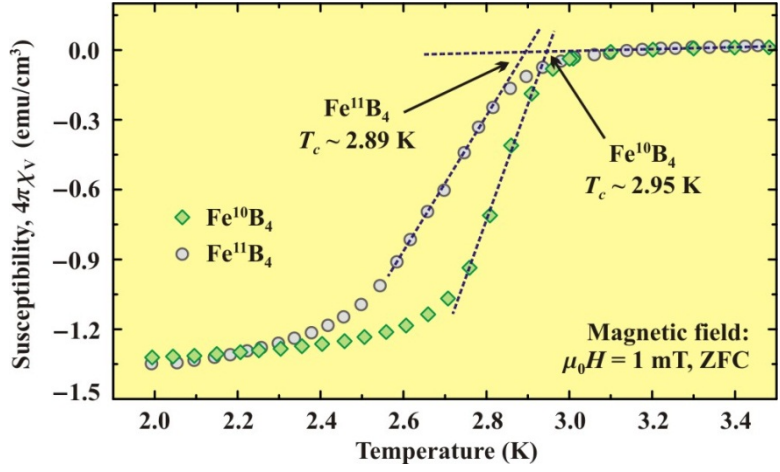

FIG. 2 (color online). Magnetic susceptibility of FeB$_4$ measured in an applied field of 1 mT after zero-field cooling (ZFC). The susceptibility is normalized to the unit of volume ($\chi_V$) and multiplied by $4\pi$ to facilitate the comparison with the expected value of $4\pi\chi_V = -1$ for the ideal bulk superconductor. Two sets of data were collected on the samples enriched with $^{10}$B and $^{11}$B isotopes. Dashed lines denote the procedure for determining $T_c$ (see Ref. [20]).

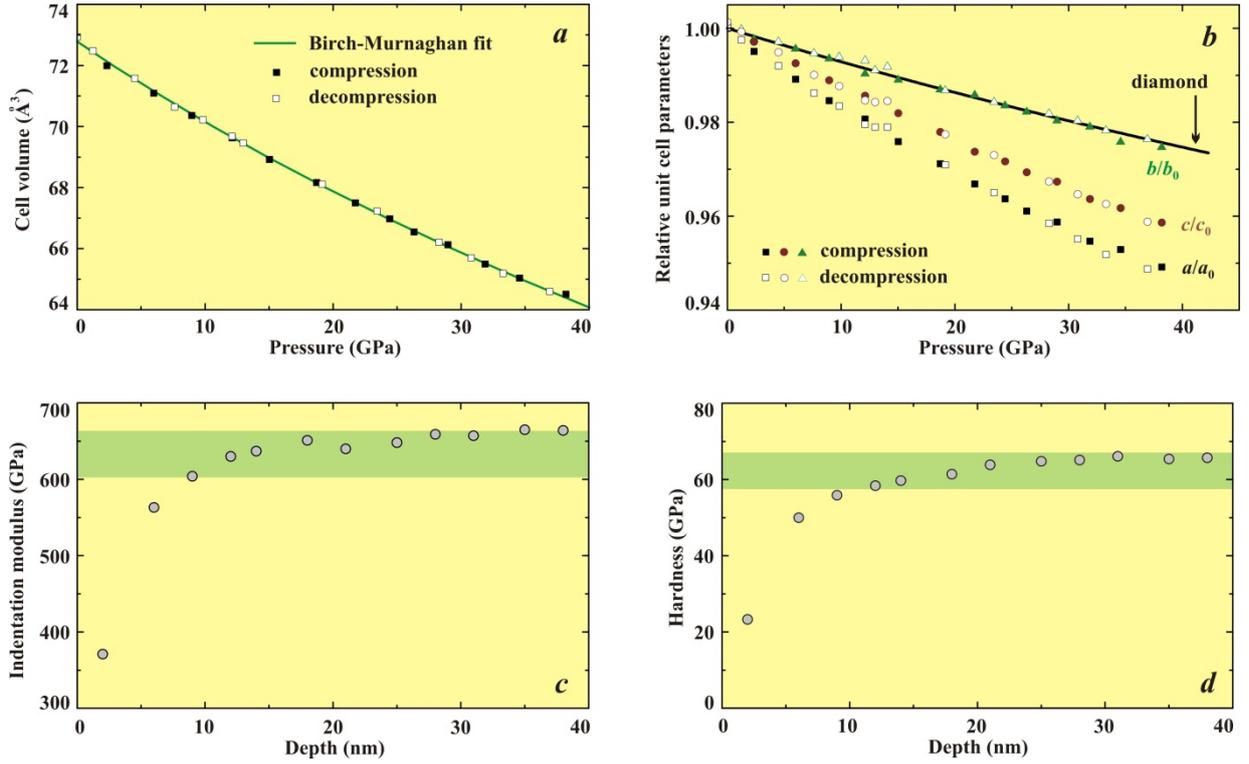

FIG. 3 (color online): Compressibility of FeB$_4$ and the results of nanoindentation measurements. (*a*) The pressure dependence of the unit cell volume based on single crystal X-ray diffraction data. The fit of the pressure-volume data with the third-order Birch-Murnaghan equation of state (solid line) gave the bulk modulus $K$ = 252(5) GPa, $K´$ =3.5(3), and $V_0$ =72.79(4) Å$^3$/unit cell. (*b*) The relative changes of the unit cell parameters as a function of pressure. The stiffness of the FeB$_4$ structure along the *b*-direction is the same as that of diamond (continues line according to Ref. [25]). Filled symbols represent the data points obtained on compression and open ones – on decompression. The uncertainties are not shown since they are smaller than the size of symbols in the figure. (*c*) Depth dependent average values of indentation modulus. (*d*) Hardness of FeB$_4$. Load-displacement curves without pop-ins have been used for evaluation with tip compression correction.